\documentclass[conference]{IEEEtran}
\IEEEoverridecommandlockouts                              

\usepackage{amsfonts}

\usepackage{cite}
\usepackage{subfigure}
\usepackage{amsmath}
\usepackage{booktabs}
\usepackage{url}
\usepackage{array}
\usepackage[dvips]{graphicx}
\usepackage{color}
\usepackage{amssymb}
\usepackage{psfrag}
\usepackage{listings}
\usepackage{caption}

\newcommand{\ignore}[1]{}
\IEEEoverridecommandlockouts

\newcommand{\DPoint}[1]{}

\definecolor{Maroon}{rgb}{0.5,0,0}
\definecolor{DarkOliveGreen}{rgb}{0,0.5,0}
\lstdefinelanguage{XML}
{
  basicstyle=\ttfamily\footnotesize,
  morestring=[b]",
  moredelim=[s][\bfseries\color{Maroon}]{<}{\ },
  moredelim=[s][\bfseries\color{Maroon}]{</}{>},
  moredelim=[l][\bfseries\color{Maroon}]{/>},
  moredelim=[l][\bfseries\color{Maroon}]{>},
  morecomment=[s]{<?}{?>},
  morecomment=[s]{<!--}{-->},
  commentstyle=\color{DarkOliveGreen},
  stringstyle=\color{blue},
  identifierstyle=\color{red}
}

\begin{document}

\title{Utility of Traffic Information in Dynamic Routing: Is Sharing Information Always Useful?}

\author{
\authorblockN{Mohammad Shaqfeh$^*$, Salah Hessien$^{\dagger}$ and Erchin Serpedin$^{\ddagger}$} \\ 
\authorblockA{$^*$Department of Electrical and Computer Engineering,
Texas A\&M University at Qatar, Doha, Qatar. \\
 Email:  Mohammad.Shaqfeh@qatar.tamu.edu} \\

\authorblockA{$^{\dagger}$Department of Electrical and Computer Engineering,
McMaster University, Hamilton, Ontario, Canada. \\
 Email:  salahga@mcmaster.ca} \\

\authorblockA{$^*$Electrical and Computer Engineering Dept.,
Texas A\&M University, College Station, TX 77843-3128. \\
 Email:  eserpedin@tamu.edu} \\
}

\maketitle

\begin{abstract}
Real-time traffic information can be utilized to enhance the efficiency of transportation networks
by dynamically updating routing plans to mitigate traffic jams. 
However, an interesting question is whether the network coordinator
should broadcast real-time traffic information to all
network users or communicate it selectively to some of them.
Which option enhances the network efficiency more?

In this work, we demonstrate using simulation experiments that 
sharing real-time traffic information with all network users is sub-optimal,
and it is actually better to share the information with a majority subset of the total population in order to improve the overall
network performance. This result is valid under the assumption that each network user decides it's route to destination
locally.
\end{abstract}


\section{Introduction}
\label{sec:introduction} 

With the advancement of sensing, surveillance, wireless connectivity and Internet-of-Things (IoT) technologies,
current transportation networks can be supported with reliable sources of information about the traffic flow in the network
and the congestion levels all over the network.
Traffic information is a good asset that can be utilized by both the network operator and users to enhance the efficiency
of traffic flow. 
In particular, we can think of two main domains of exploiting traffic information. The first one is the use of recorded
historic information about traffic flows in machine learning applications such as 
traffic prediction and forecasting, and trips' scheduling and planning.
The second one is the use of real-time (i.e., live) information about the network congestion levels in adaptive control applications
such as updating routing plans dynamically to arrive at destination as quick as possible. 
In this work, we focus on the utility of real-time
information in dynamic routing to enhance the efficiency of traffic flow and mitigate traffic jams.
 
The advantages of utilizing traffic information have been demonstrated in a number of papers in the literature.
For example, in \cite{Kim}, the value of real-time traffic information from the perspectives of a time-sensitive freight service
that aims to determine the optimal routing policies as well as departure times is examined. The paper demonstrated the advantages of real-time information in terms of total cost savings and vehicle usage reduction.
In another work, it is demonstrated that even when information is shared in a decentralized manner 
it can considerably improve the network performance \cite{Claes}. Furthermore, a decentralized approach to 
collect and distribute traffic forecast information is proposed as an alternative to depending solely 
on a central network coordinator.

There are other works that focused mainly on developing routing algorithms.
For example, a nice survey of search algorithms that can be used for routing problems 
is provided in the introduction of \cite{Xiao}.
Then, a new algorithm is also proposed to exploit the information that is collected while a vehicle 
is en route to improve the routing quality.
In \cite{Gitae}, the problem of estimating the probability distribution of travel time is investigated.
A Markov decision process model is used, and approximate dynamic programming is applied.
In another work, the combination of historical and real-time information is investigated 
in the context of a centralized navigation system \cite{Chen}. An integrated approach of
offline computation of a number of good candidate routes using historical data,
and selecting the best candidate among the pretrip identified routes using real-time data is proposed. 
The main motivation is to make real-time routing less expensive in terms of computation load.

All of these works give examples of how traffic information can be utilized by a network user to improve
it's path planning (i.e., routing) decisions. There are also some works that consider 
network perspectives and means of collecting information.
In \cite{Guo}, collecting and aggregating real-time data from road segments 
and vehicles with the aid of IoT technologies is studied.
In \cite{He}, it is proposed to employ large-scale social signal data to estimate travel demands, and to help travelers who
experience severe congestion by proposing new routes to them. 
The list of references in this quick literature review is by no means exhaustive. There are also other interesting papers
on dynamic routing and on collecting and utilizing traffic information to enhance the efficiency of transportation networks.

As a new contribution to the literature, we consider the problem of selectively sharing traffic information 
with the network users. Our aim is get more insights into the utility of traffic information as a function of
the percentage of network users who have access to real-time traffic information 
and, hence, can utilize it in dynamic routing.
We assume that the central network operator (coordinator) has access to real-time
traffic information and can communicate it in a selective manner to the vehicles in the network.
The network operator has the objective of improving the overall network performance.

To study this problem, we conducted a series of experiments using our developed 
large-scale traffic simulator (SmarTTS \cite{SmarTTS}).
As an interesting and insightful observation,
we demonstrate that sharing real-time information with all network users is sub-optimal,
and it is actually better to share the information with only 50\% - 70\% of the total population to improve the overall
network performance. This result is valid under the assumption that each network user decides it's route to destination
locally.

The rest of this paper is organized as follows. In Section~\ref{sec:setup}, 
we give an overview of the investigated research problem and the used simulation setup to study it.
Then, in Section~\ref{sec:eval}, we discuss the main results, observations and key insights of the 
simulation experiments.
Finally, we give some conclusions and suggested future work directions.


\section{Problem Statement and Experiment Setup}
\label{sec:setup} 

\subsection{Problem Definition}
\label{sec:problem}

We assume a smart transportation network environment that has access to real-time information about the
traffic flow states all over the network via sensing and surveillance technologies and/or via communication with the 
vehicles in the roads. Additionally, we assume that the smart environment has the ability to communicate with the network users (i.e., vehicles in the roads) to support them with traffic information, if needed.
The network users can get benefit from the communicated traffic information by updating their travel plans (i.e., routing decisions)
locally (i.e., by themselves) to avoid traffic congestion and arrive at their destinations in the shortest possible duration.

We distinguish between three possible scenarios in terms of sharing the traffic information:
\begin{itemize}
\item Scenario (A): No traffic information is shared at all. We assume in this case that the 
network users rely solely on the static geographical
map of the network, and choose the routes of the {\bf shortest distance} to their intended destinations. We label
this scenario as {\it static routing}.
\item Scenario (B): Real-time traffic information is updated regularly 
over short periods of time and communicated to \underline{all} users
of the network. We assume in this case that the users dynamically update their travel routes to follow the path with
the {\bf shortest expected travel time}. We label this scenario as {\it dynamic routing}.
\item Scenario (C): The smart environment \underline{selectively} communicates 
the real-time traffic information to a subset of the total
population of the vehicles in the network. So, in this case, some network users will use {\it dynamic routing}
to achieve the shortest expected travel time,
and the others will use {\it static routing} following the shortest distance to destination.
\end{itemize}

Our objective is to compare the prospected performance of these three scenarios, and to get insights about the situations
in which one of them becomes clearly favorable. Our performance metric is the efficiency of traffic flow all over the network.
So, we choose the average travel time of all users as the performance metric that we aim to minimize.
Enhancing traffic efficiency from the network perspective will certainly be reflected on enhancing the performance
of the individual network users as well.

We use our developed large-scale traffic simulator platform, called SmarTTS \cite{SmarTTS}, to run a series of experiments
of the three scenarios that we want to investigate. 
The SmarTTS simulator has a number of features that are useful for our purpose.
It enables us to generate network map freely, specify traffic injection rate and distribution, 
adjust traffic control operations (speed limits, traffic light operation, etc.) as needed,
specify the information that can be communicated from the infrastructure to the vehicles,
and generate detailed simulation trace file that can be used for statistical analysis.
Furthermore, several vehicle mobility models, such as Intelligent Driver Model (IDM) and Krauss Car-Following Model (CFM) \cite{Kanagaraj}, are implemented in SmarTTS to mimic realistic traffic mobility. 
Those models deal with the vehicles propagation along the road.
The lateral behavior of the vehicle in a multi-lane road is modeled using the MOBIL algorithm for lane changing \cite{Kesting}.
Furthermore, a number of models are implemented to support collision avoidance behavior
\cite{Kiefer, Knipling, NHTSA}.

\subsection{Experiment Setup}

\begin{figure}
\centering
\includegraphics[width=\linewidth]{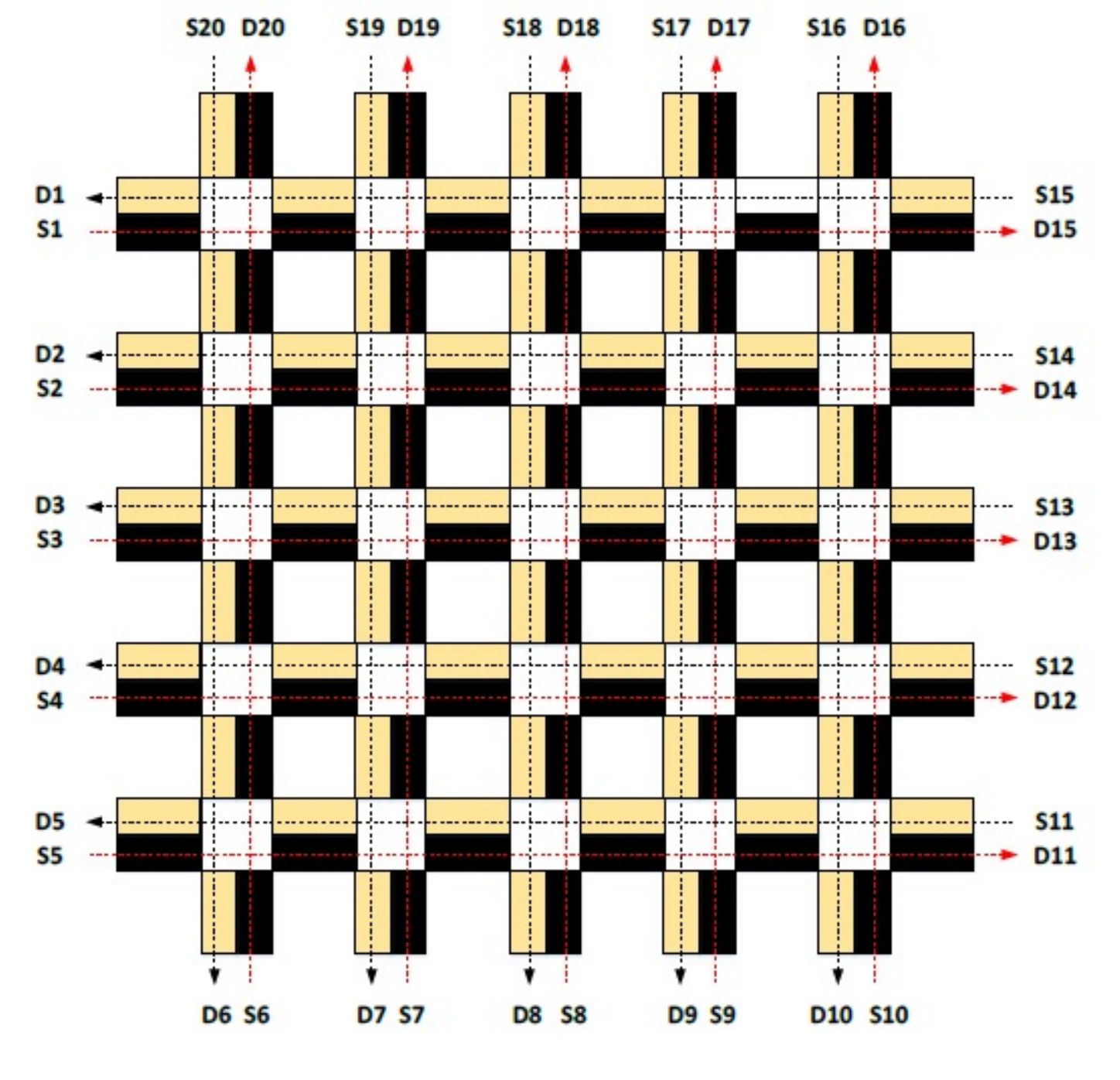}
\caption{Simulated traffic network map.}
\label{fig:Layout}
\end{figure}

A hypothetical Manhattan-like traffic network was simulated assuming non-uniform traffic distribution.
The network consists of five horizontal (i.e., east-west) highways and five vertical (i.e., north-south) highways,
as shown in Fig.~\ref{fig:Layout}. Each highway is assumed to have five lanes, and the network spans an
area of 12 km by 12 km. All 25 intersections of the network are controlled by traffic lights that are actuated smartly
based on the relative traffic volume in each direction of the intersection.

For the purpose of our study, new traffic are continuously injected to the network 
from the entry roads on the boarders of the simulation area,
which are labeled as $S_1$, $S_2$, ..., $S_{20}$ in Fig.~\ref{fig:Layout}. These are the traffic sources.
Similarly, the traffic destinations are the outward directed roads at the boarders of the map, and they are
labeled as $D_1$, $D_2$, ..., $D_{20}$. 
Traffic generation at the sources follows a Poisson process probability distribution, with traffic generation density
$\lambda_{S_i}$ (in vehicles per hour) of source $S_i$ adjusted as a parameter of the simulation to control
the congestion level in our experiments. The average (of all traffic sources) traffic generation density is denoted as $\lambda$
($\lambda=\frac{1}{20}\sum_{i=1}^{20}\lambda_{S_i}$). 
To mimic real-traffic scenarios, we assume that the generation density of the sources varies such that traffic
is injected form some entries more heavily than in other entries. 

Similarly, we control, as a simulation parameter, the traffic distribution matrix of the traffic flow from each source to each destination.
For example, we can make a certain percentage of the vehicles injected from source $S_i$ go to destination $D_j$.
We have generated a non-uniform distribution matrix for the purpose of our experiments.
For example, we assumed that the traffic that is injected from $S_1$ is distributed as follows; 15\% to $D_5$,
20\% to $D_8$, 10\% to $D_{10}$, 10\% to $D_{12}$, 10\% to $D_{14}$, 15\% to $D_{15}$,
10\% to $D_{17}$, and 10\% to $D_{18}$.

We simulate a two-hour operation of the transportation network, and measure the average total trip time 
(from source to destination), denoted $\tau_\text{avg}$ of all vehicles in the network.
Travel times in the last 70 minutes of the simulation time are considered in the evaluation of the performance metric in order
to make sure that the network has already reached steady state.
We test the three possible routing scenarios, that were introduced in Section~\ref{sec:problem}.
Whenever the experiment is repeated to test another scenario, we use the same traffic distribution 
matrix to make a valid comparison between the three cases.

\subsection{Routing Algorithms}

The routing decisions, whether they are based on shortest distance or shortest expected travel time,
can be obtained by modeling the transportation network as a directed graph $G$.
The graph is made up of two objects, a set $V$ whose elements are called “vertices” or “nodes” which represent
all road segments in the network; and a set $E$ whose elements are called “edges”. Each edge represents a
connection between two road segments.  
Each edge is assigned a weight $W$, which represents a cost metric for going 
from one road segment to another one that is connected 
to it at an intersection. The graph is commonly denoted as $G(V, E)$.

The problem of optimal routing can be formulated as finding the shortest path (i.e., with least accumulated edges' weights)
between two nodes in the graph (i.e., the source and destination node).
This problem can be solved using a number of well-known search algorithms such as the {\it Dijkstra Algorithm} \cite{Dijkstra},
which we have implemented and used for our simulations.

In static routing the weight of a certain edge corresponds to the travel distance between its two
nodes, which is a static weight that is based on the geographic network map.
On the other hand, dynamic routing calculates the weights of the edges based on the moving average
time that the most recent vehicles have spent to travel between the two
 vertices (i.e., the time to go from the starting point of one road segment
to the starting point of the next road segment including the expected waiting time at the intersection between them). 
This is a dynamic metric that changes 
over time based on the traffic congestion state on each road segments. 
To simulate the dynamic routing case, we assume that these weights get
periodically updated every $\tau_\text{up}$ minutes, which is a simulation parameter. 

So, in summary, the cost metric (i.e. edges weights) in each scheme is calculated based on the
following equation,
\begin{equation*}
W(v_i, v_j) = \left\{
\begin{matrix}
D(v_i, v_j) & \text{Static routing} \\
\frac{1}{N_\text{car}}\sum_{k=1}^{N_\text{car}}\tau_k(v_i, v_j) & \text{Dynamic routing}
\end{matrix}
\right.
\end{equation*}
where $D(v_i, v_j)$ is the travel distance between vertex $v_i$ and $v_j$, and $\tau_k(v_i, v_j)$ is the travel time
that the $k$th last vehicle has taken during the travel between the two vertices. 
In this case, only the last vehicles that moved between these
two vertices are used in computing the moving average, where $N_\text{car}$ is the number of vehicles that is 
used in calculating the moving average.

\section{Simulation Results And Discussions}
\label{sec:eval}

\begin{figure}
\centering
\includegraphics[width=\linewidth]{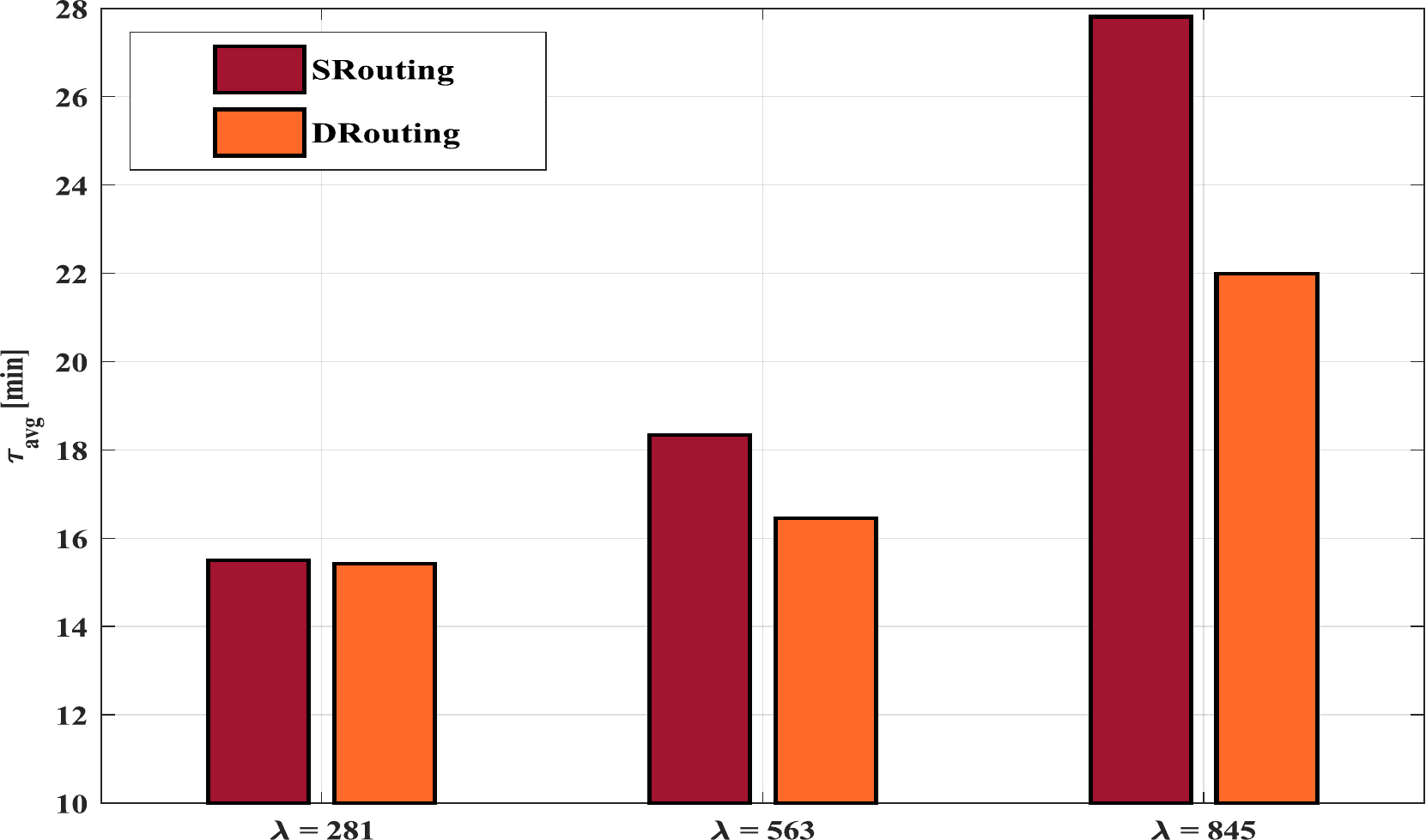}
\caption{Comparison of static routing and dynamic routing in terms of the total average travel time of all vehicles in the network
over 70 minutes simulation time.}
\label{fig:fifth}
\end{figure}

\begin{figure}
\centering
\includegraphics[width=\linewidth]{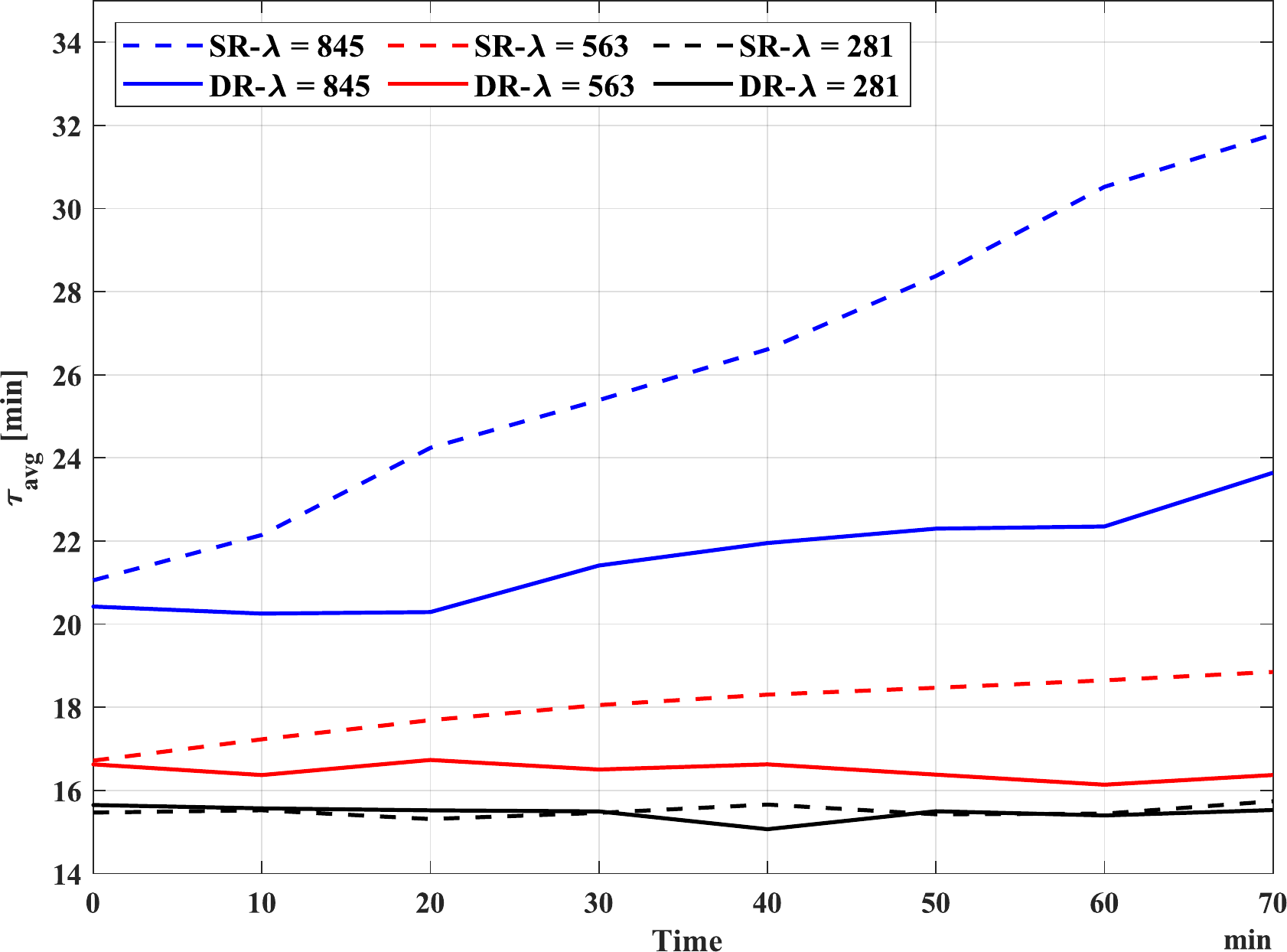}
\caption{Comparison of static routing and dynamic routing in terms of the moving average travel time of the vehicles in the network.}
\label{fig:second}
\end{figure}

It is intuitive to predict that dynamic routing should be better than static routing since it exploits real-time traffic information 
to avoid congested roads. This is confirmed in Fig.~\ref{fig:fifth}, which compares static and dynamic routing for 
three different values of the average traffic generation density $\lambda$ (in vehicles per hour per source). Since we have 20 sources in our network, the total generated traffic rate in the network is 20$\times\lambda$.
The results show that dynamic routing is always advantageous compared to static routing.
However, the advantage becomes minimal when the traffic volume is low (such as when $\lambda = 281$).
Furthermore, as shown in Fig.~\ref{fig:second}, when the traffic volume is high, 
dynamic routing does not only save travel time, but it does also keep the network stable, since in the case of static routing
the moving average travel time keeps growing up.
So, sharing real-time traffic information is vital in such cases.

\begin{figure}
\centering
\includegraphics[width=\linewidth]{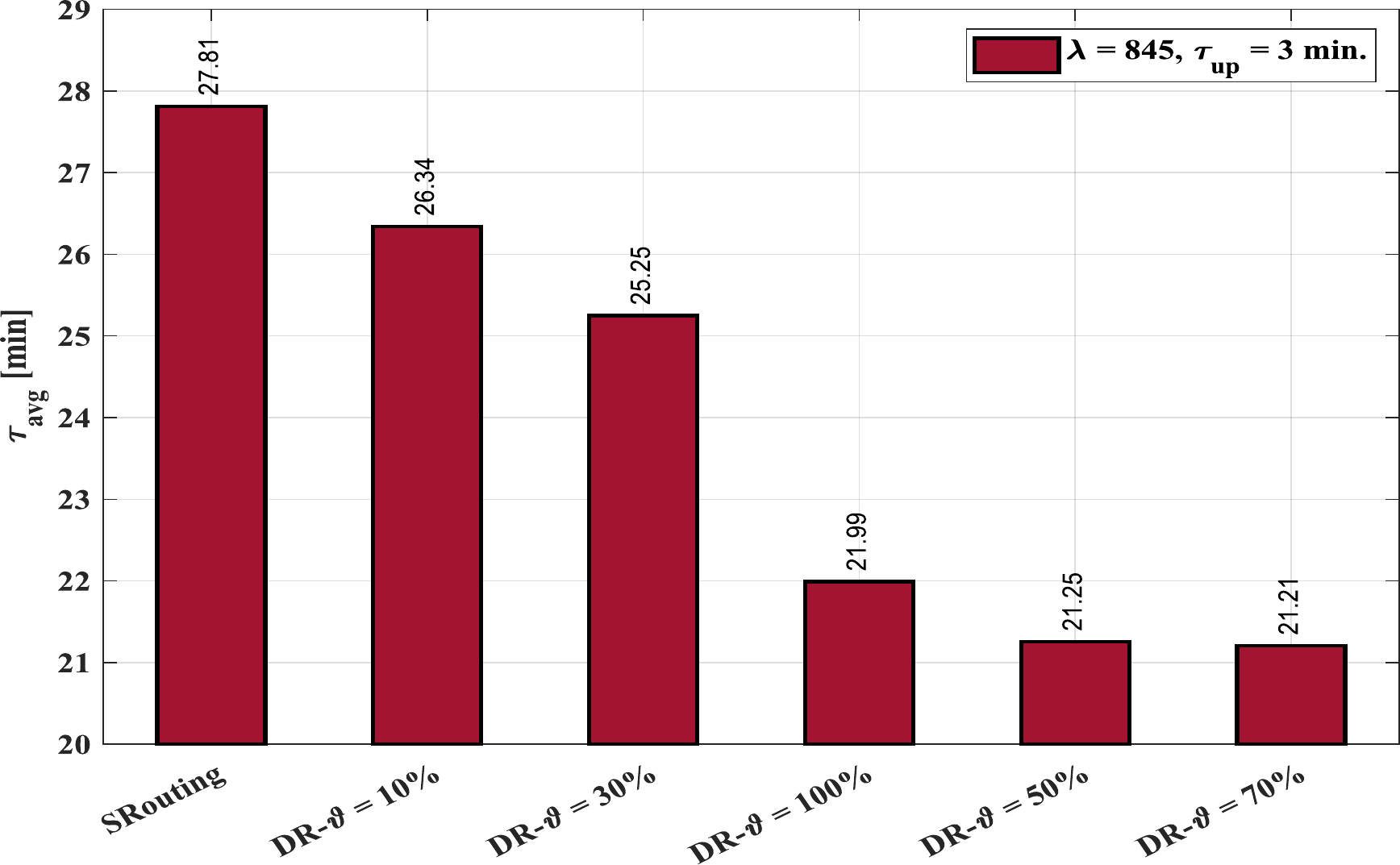}
\caption{Comparison of static routing, dynamic routing and selective dynamic routing, with only a percentage $\vartheta$ 
of the total number of vehicles follow dynamic routing.}
\label{fig:sixth}
\end{figure}

The more interesting and insightful results are shown next in Fig.~\ref{fig:sixth} which shows the results
including the scenario when only a subset of the network users (i.e., vehicles) receive real-time traffic information and 
utilize it in dynamic routing, and the remaining vehicles stick to shortest distance static routing.
Here we assume that a percentage $\vartheta$ of the total number of network users are selected randomly
to receive real-time traffic information. 
In this case, the overall network performance does not only improve upon the static routing scenario, but it does also
improve upon the case when all vehicles perform dynamic routing. This is true given that $\vartheta$ is selected properly.
For example, when only half of the total number of vehicles receive real-time traffic information, the overall 
performance is better than when all vehicles receive real-time traffic information. It is slightly further 
improved when only 70\% of the overall population receive real-time traffic updates.
This is an interesting and counter-intuitive observation because it means that the utility of information is not
monotonically increasing with the number of users who receive the information and utilize it locally.
One might expect that static routing gives a lower bound of the performance and dynamic routing gives
an upper bound, and having a mix of them will give a performance that lies between these two bounds. 
However, this is not true at all. Sharing information with all users such that all of them follow dynamic routing is not an upper
bound.

To explain these results, we should take into consideration that we assume that all routing decisions are localized at
each individual vehicle, and each one of them will seek its own advantage. 
So, sharing information about real-time traffic will make all users in nearby of congested areas update their plans simultaneously,
and hence the traffic congestion will eventually move from one place to another. 
On the other hand, the ignorance of some vehicles helps to achieve a more even distribution of the traffic in the network, which resolves the occurring traffic jams.
It is intuitive to predict that some kind of centralized coordination, that learns the intended destinations of 
all users and runs a joint routing problem for all of them, would be advantageous at both individual vehicle level and at the overall network level. However, this would be a computationally expensive task that can be successfully replaced by the simpler strategy 
of just communicating with a subset of the users, and let them update their routes locally.

\begin{figure*}
\centering
\includegraphics[width=\linewidth]{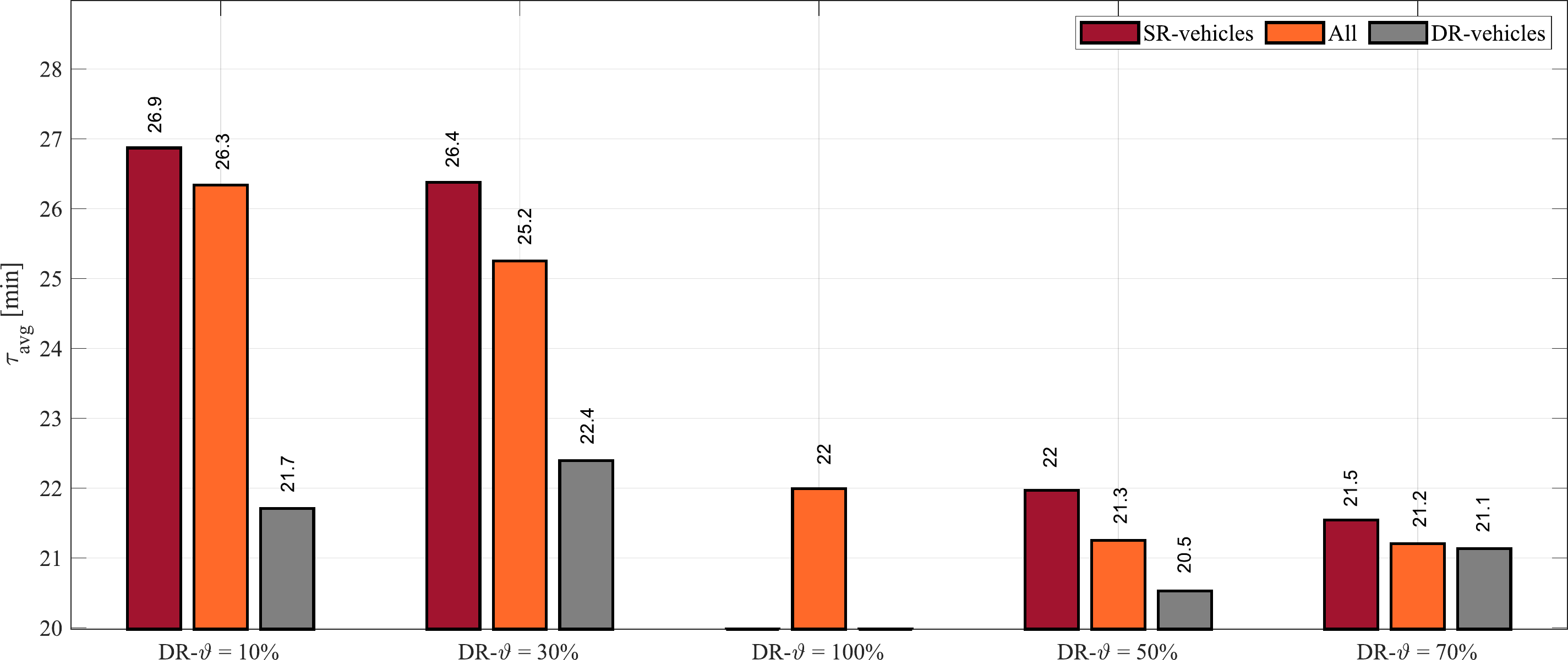}
\caption{Comparison between static routing users and dynamic routing users in scenarios when a percentage $\vartheta$
of users who follow dynamic routing.}
\label{fig:fourth}
\end{figure*}

To get further insight, we differentiate in Fig.~\ref{fig:fourth} between the static routing users and the dynamic routing users
in terms of total average travel time of all vehicles in the same class (i.e., static or dynamic routing). Also, the overall average 
is also shown for comparison. As expected, the performance of the dynamic routing users are better than the static routing users.
However, it should be noted than when $\vartheta$ is 50\% or 70\%, the performance of the static-routing users
does not get worse than the case when all users follow dynamic routing (i.e., $\vartheta$ is 100\%). It is only that the dynamic routing users get some extra advantage in this case relative to the all-dynamic routing scenario.  
It is as if the “dynamic-routing” vehicles get an extra benefit due to the “ignorance”' of the “static-routing” vehicles in this case.
If all users receive real-time traffic information, then no one will receive extra performance gains.

This result raises fairness concerns. If the central network coordinator wants to communicate the traffic information to a subset
of the total network population with the sole goal of improving the overall network performance, some users will receive 
a preferential treatment. So, selecting the users to receive traffic information can be selected randomly so that everyone has
the same chances of getting such preferential treatment. 
Having said that, it should be noted that in the case when $\vartheta=70\%$, the differences between the static-routing 
and dynamic-routing users become very minimal.

\begin{figure*}
\centering
\includegraphics[width=\linewidth]{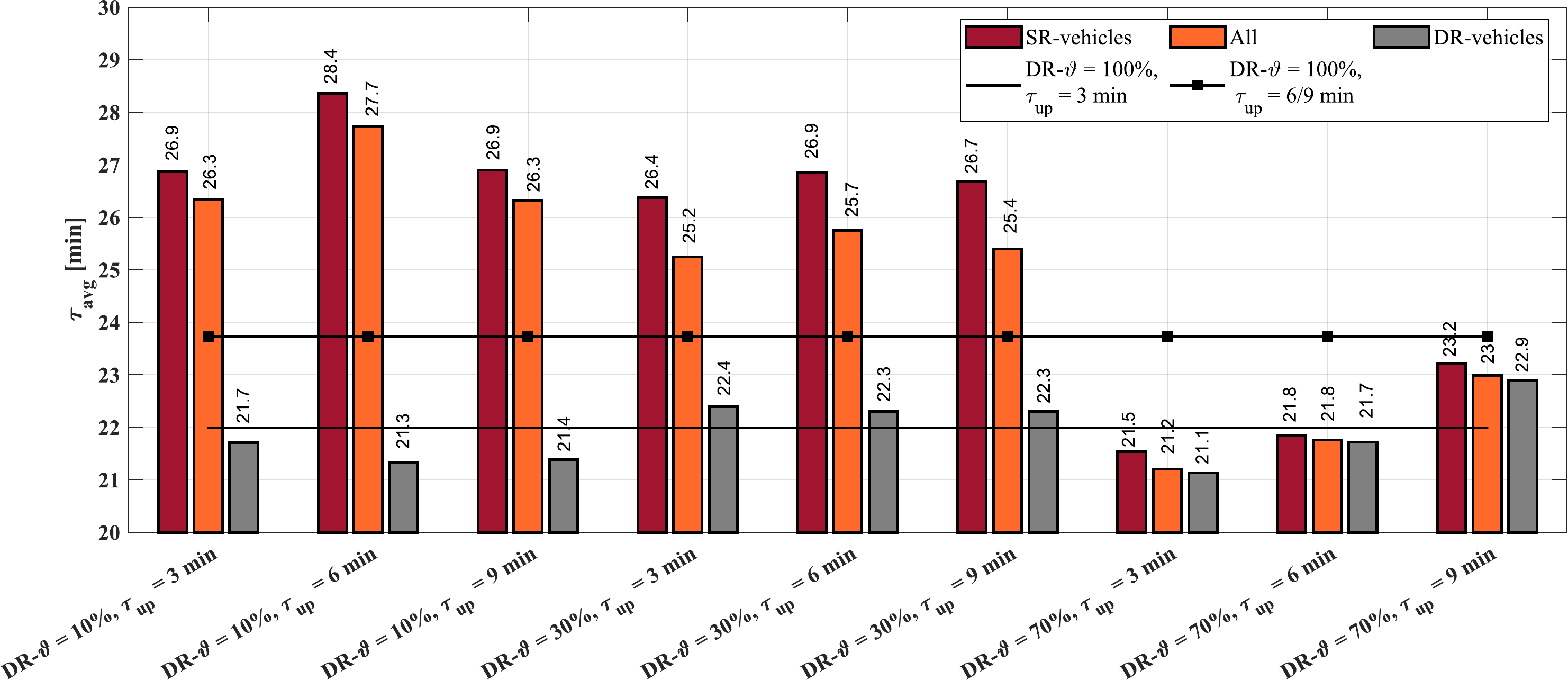}
\caption{Performance of the network users for different update rate $\tau_\text{up}$ of the network real-time information.}
\label{fig:third}
\end{figure*}

Finally, Fig.~\ref{fig:third} show the results for different traffic information update rate $\tau_\text{up}$, of 3, 6 and 9 minutes.
The timing between the real-time traffic information updates affect the frequency by which the travel plans get re-calculated
in dynamic routing. The results show that having more frequent information updates is advantageous 
for the cases when $\vartheta=70\%$ and $\vartheta=100\%$ (i.e., all users follow dynamic routing).
This is not the case for $\vartheta=10\%$ and $\vartheta=30\%$.

\section{Conclusion and Future Work}
\label{sec:conclusion}
The key message of this work is that sharing real-time traffic information with all 
users of a transportation network is sub-optimal, and it is better to be selective in sharing traffic information 
in order to enhance the network efficiency.
We have demonstrated this result via a number of simulation experiments of a large-network map.
In our experiments, we assumed that traffic information are utilized locally at each vehicle to update
its routing decisions dynamically.

The results of this work can be further elaborated by supporting them with some theoretical foundations.
In particular, we can establish a link between these observed experimental results 
and the concepts of ``Nash' equilibrium''and ``Pareto-Optimality'' that are 
discussed in game theory (e.g., \cite{Game}).
We can think of the operating point of the network to be achieving a Nash equilibrium since
each user would seek to achieve it's best performance, and this is far from the Pareto-optimal operating points
of the network. However, the role of the network coordinator will be to selectively share information
such that the Nash equilibrium becomes closer to Pareto-optimality. This theoretical
framework will be elaborated in an extension of this work.
Furthermore, selecting the subset of users to receive traffic information can be further investigated.
It may be more efficient to choose the users based on their locations instead of communicating with the users
all over the network in a completely random way.
Moreover, comparing the suggested simple selective information sharing scheme
with collaborative routing strategies, which are computationally expensive, would be a good continuation of this work.   

Finally, the shown results are relevant not only for transportation networks, but also 
for other network applications as well. This is because 
sharing information over networks is a fundamental problem that is applicable to many disciplines.


\bibliographystyle{IEEEbib}
\bibliography{./BIBFILES/papers,./BIBFILES/books}

\end{document}